\documentclass[aps,floats]{revtex4}
\usepackage{amsmath,amssymb}
\usepackage{graphicx,epsfig}

\begin{document}
\bibliographystyle {plain}

\def\oppropto{\mathop{\propto}} 
\def\opsimeq{\mathop{\simeq}}
\def\opoverderline{\mathop{\overline}}
\def\operarrow{\mathop{\longrightarrow}}
\def\opsim{\mathop{\sim}}

\def\fig#1#2{\includegraphics[height=#1]{#2}}
\def\figx#1#2{\includegraphics[width=#1]{#2}}


\title{Pure and Random Quantum Ising Chain :
\\ Shannon and R\'enyi entropies of the ground state via real space renormalization   } 


 \author{ C\'ecile Monthus}
  \affiliation{Institut de Physique Th\'{e}orique, 
  CNRS and CEA Saclay, 
  91191 Gif-sur-Yvette cedex, France}

\begin{abstract}
The Shannon and the R\'enyi entropies of the ground state wavefunction in the pure and in the random quantum Ising chain are studied via the self-dual Fernandez-Pacheco real-space renormalization procedure. In particular, we analyze the critical behavior of the leading extensive term at the quantum phase transition : the derivative with respect to the control parameter is found to be logarithmically divergent in the pure case, and to display a cusp singularity in the random case. This cusp singularity for the random case is also derived via the Strong Disorder Renormalization approach.

\end{abstract}

\maketitle

\section{ Introduction} 

After its introduction in fluid dynamics in order to
characterize the statistical properties of turbulence 
(see the book \cite{frisch} and references therein),
the notion of multifractality has turned out to be relevant in many areas of physics
(see for instance \cite{halsey,Pal_Vul,Stan_Mea,Aha,Meakin,harte,duplantier_houches}),
in particular at critical points of disordered models
like Anderson localization transitions \cite{janssenrevue,mirlinrevue}
or in random classical spin models
\cite{Ludwig,Jac_Car,Ols_You,Cha_Ber,Cha_Ber_Sh,PCBI,BCrevue,Thi_Hil}.
More recently, the wavefunctions of manybody quantum systems 
have been found to be generically multifractal, with many studies
concerning the Shannon-R\'enyi entropies of the ground states of pure quantum spin models
\cite{jms2009,jms2010,jms2011,moore,grassberger,atas_short,atas_long,luitz_short,luitz_o3,luitz_spectro,luitz_qmc,jms2014,alcaraz}. The understanding of Shannon-R\'enyi entropies of excited states
in disordered models is also essential to characterize the Many-Body Localization transition 
 \cite{luca_mbl,luitz_mbl}.

In the present paper, we analyze via real-space renormalization
 the multifractal properties
of the ground state of the quantum Ising chain 
\begin{eqnarray}
H=- \sum_{i}  J_i \sigma_{i}^z \sigma_{i+1}^z - \sum_{i}  h_i \sigma_{i}^x
\label{h1d}
\end{eqnarray}
where the ferromagnetic couplings $J_i>0$  and the transverse fields $h_i>0$ 
are either uniform or random.
For the pure chain, we compare with the previous results of Refs \cite{jms2009,jms2010,moore,grassberger,atas_short,atas_long,luitz_spectro,jms2014,alcaraz} in order to test the validity of the RG approach. For the random case, we are not aware of previous works concerning the multifractality of the ground state.

The paper is organized as follows.
In section \ref{sec_multif}, we recall the multifractal formalism for quantum wavefunctions.
In section \ref{sec_rg}, we derive the real space 
renormalization rule for the Shannon-R\'enyi entropies.
The application to the pure and to the random quantum Ising chain
are described in sections \ref{sec_pure} and \ref{sec_random} respectively.
Our conclusions are summarized in section \ref{sec_conclusion}. 

\section{ Reminder on multifractality of manybody wavefunctions }

\label{sec_multif}

For an Hilbert space of size $M$ (growing exponentially with the volume),
the expansion of a given wave-function $\vert \psi >$ onto
a basis of M vectors $\vert m > $
\begin{eqnarray}
\vert \psi > = \sum_{m=1}^M \psi_m \vert m> 
\label{exppsi}
\end{eqnarray}
involves $M$ coefficients $\psi_m $ subjected to the global normalization
\begin{eqnarray}
1= <\psi \vert \psi > = \sum_{m=1}^M \vert \psi_m \vert^2
\label{normaexppsi}
\end{eqnarray}
The multifractal formalism allows to characterize
 the statistical properties of these $M$ weights
$ \vert \psi_m \vert^2 $ as we now recall.

\subsection{ Inverse Participation Ratios and  Shannon-R\'enyi entropies }
  
The Inverse Participation Ratios $Y_q(M) $ define the exponents $\tau(q)$
\begin{eqnarray}
Y_q(M) \equiv  \sum_{m=1}^M \vert \psi_m \vert^{2q} \oppropto_{M \to +\infty}  M^{- \tau(q)}
\label{IPR}
\end{eqnarray}
The related R\'enyi entropies involve the generalized fractal dimensions $D_q$
 \begin{eqnarray}
S_q(M) \equiv  \frac{ \ln Y_q(M) }{1-q} \oppropto_{M \to +\infty} D(q) \ln M
\label{renyi}
\end{eqnarray}
with the simple relation
 \begin{eqnarray}
D(q) = \frac{ \tau(q)}{q-1 }
\label{Dqtauq}
\end{eqnarray}
For $q=1$, Eq. \ref{renyi} corresponds to the standard Shannon entropy
 \begin{eqnarray}
S_1(M) \equiv - \sum_{m=1}^M \vert \psi_m \vert^{2} \ln \vert \psi_m \vert^{2}
\oppropto_{M \to +\infty} D(1) \ln M
\label{shannon}
\end{eqnarray}

\subsection{ Multifractal spectrum $f(\alpha)$ }

\label{sec_falpha}

Among the $M$ configurations, the number ${\cal N}_M(\alpha)$ of configurations $m$
having a weight of order $\vert \psi_m \vert^2 \propto M^{-\alpha}$
defines the multifractal spectrum $f(\alpha)$
\begin{eqnarray}
{\cal N}_M(\alpha) \oppropto_{M \to \infty} M^{f(\alpha)}
\label{nlalpha}
\end{eqnarray}
The saddle-point calculus in $\alpha$ of the Inverse Participation Ratios of Eq. \ref{IPR}
yields
\begin{equation}
Y_q(M) \simeq \int d\alpha \ M^{f(\alpha)} \ M^{- q \alpha} \oppropto M^{f(\alpha_q)- q \alpha_q } = M^{-\tau(q)}
\label{saddle}
\end{equation}
where the saddle point value $\alpha_q$ is determined by the condition
\begin{eqnarray}
   f'(\alpha_q) && = q  
\label{legendrec}
\end{eqnarray}
i.e. $\tau(q)$ and $f(\alpha)$ are related via the Legendre transform
\begin{eqnarray}
   \tau(q)+f(\alpha) && = q \alpha 
\nonumber \\
   \tau'(q) && =  \alpha 
\nonumber \\
f'(\alpha) && = q 
\label{legendre}
\end{eqnarray}

This yields the following parametric representation of $f(\alpha)$
\begin{eqnarray}
   \alpha_q =\tau'(q)= \frac{-\partial_q \ln Y_q}{\ln M } = - \frac{\sum_{m=1}^M \vert \psi_m \vert^{2q} \ln \vert \psi_m \vert^{2}}
{\sum_{m=1}^M \vert \psi_m \vert^{2q} \ln M }  
\label{alphaq}
\end{eqnarray}
\begin{eqnarray}
   f(\alpha_q) = - \frac{\sum_{m=1}^M \vert \psi_m \vert^{2q}
 \ln \frac{ \vert \psi_m \vert^{2q}}{\sum_{m=1}^M \vert \psi_m \vert^{2q}}}
{\sum_{m=1}^M \vert \psi_m \vert^{2q} \ln M }  
\label{falphaq}
\end{eqnarray}

Let us now mention some important special values of the index $q$.

For $q=0$, the IPR of Eq. \ref{IPR} simply measures the size of the Hilbert space $Y_0(M)=M $
corresponding to $\tau(0)=-1$ and $D(0)=1$ and to $f(\alpha_0)=1$ :
 this means that there exists an extensive
number $O(M)$ of configurations having a weight $M^{-\alpha_0}$,
so that 
\begin{eqnarray}
   \alpha_0 = - \frac{\sum_{m=1}^M  \ln \vert \psi_m \vert^{2}} { M \ln M }  
\label{alphazero}
\end{eqnarray}
 is called the typical exponent.

For the Shannon value $q=1$, 
the IPR of Eq. \ref{IPR} is normalized to unity, so that $\tau(1)=0$
and 
\begin{eqnarray}
   \alpha_1 = - \frac{\sum_{m=1}^M \vert \psi_m \vert^{2} \ln \vert \psi_m \vert^{2}}
{ \ln M }  = f(\alpha_1) = D(1)
\label{alphaun}
\end{eqnarray}
is called the 'information dimension' that characterizes
 the Shannon entropy of Eq. \ref{shannon}. 

In the limit $q \to + \infty$, the IPR of Eq. \ref{IPR} is dominated by
the configuration having the highest weight $Y_{q \to +\infty} \simeq \vert \psi_{max} \vert^{2q} \propto M^{- q \alpha_{+\infty}} $,
 i.e. $\alpha_{+\infty}$ represents the smallest exponent
\begin{eqnarray}
   \alpha_{+\infty} = - \frac{ \ln \vert \psi_{max} \vert^{2}} { \ln M }  = D(+\infty)
\label{alphaqinfty}
\end{eqnarray}
and the corresponding singularity spectrum vanishes $f(\alpha_{+\infty})=0$

\subsection{ Application to the ground state of the quantum Ising chain   }

For the quantum Ising chain of Eq. \ref{h1d} containing $N$ spins,
the expansion of the wave function in the $\sigma^z$ basis 
\begin{eqnarray}
\vert \psi > = \sum_{S_1=\pm 1} \sum_{S_2=\pm 1} .. \sum_{S_N=\pm 1}
\psi(S_1,S_2,...,S_N) \vert S_1,S_2,..,S_N> 
\label{exppsiising}
\end{eqnarray}
involve $M=2^N$ coefficients $\psi(S_1,S_2,...,S_N)  $.
  
\subsubsection{ Ferromagnetic limit $h_i=0$ }

In the ferromagnetic limit where all transverse fields vanish $h_i=0$,  
the ground state is simply the linear combination of the two ferromagnetic states
 (both for periodic or free boundary conditions)
\begin{eqnarray}
\vert \psi >^{ferro} = \frac{ \vert +,+,..,+> +\vert -,-,..,-> }{\sqrt{2}} 
\label{exppsiferro}
\end{eqnarray}
so that among the $M=2^N$ coefficient $ \psi_m $,
only two are non-vanishing and equal to $1/\sqrt{2}$.
As a consequence the IPR of Eq. \ref{IPR} reduces to
\begin{eqnarray}
Y^{ferro}_q(M=2^N)  = 2^{1-q}
\label{IPRferro}
\end{eqnarray}
and the corresponding R\'enyi entropies all take the simple finite value
 \begin{eqnarray}
S_q^{ferro}(M) =  \frac{ \ln Y_q(M) }{1-q} = \ln 2
\label{renyiferro}
\end{eqnarray}
i.e. all generalized fractal dimensions of Eq. \ref{renyi} vanish
\begin{eqnarray}
D_q^{ferro} =0
\label{dqferro}
\end{eqnarray}

\subsubsection{ Paramagnetic limit $J_i=0$ }

In the paramagnetic limit where all ferromagnetic couplings vanish $J_i=0$,  
the ground state is simply given by the tensor product of the ground state
of $\sigma^x$ for each spin (both for periodic or free boundary conditions)
\begin{eqnarray}
\vert \psi >^{para} = \otimes_{i=1}^N \left( \frac{ \vert S_i=+> +\vert S_i=-1> }{\sqrt{2}}  \right)
\label{exppsipara}
\end{eqnarray}
i.e. all $M=2^N$ coefficients are equal to $2^{-\frac{N}{2}}$ so that
the IPR of Eq. \ref{IPR} read
\begin{eqnarray}
Y^{para}_q(M=2^N)  = 2^{N(1-q)}
\label{IPRpara}
\end{eqnarray}
and the R\'enyi entropies all have the same extensive value 
 \begin{eqnarray}
S_q^{para}(M) = - \frac{ \ln Y_q(M) }{q-1} = \ln M =N \ln 2
\label{renyipara}
\end{eqnarray}
i.e. all generalized fractal dimensions of Eq. \ref{renyi} are equal to unity
\begin{eqnarray}
D_q^{para} =1
\label{dqpara}
\end{eqnarray}

\subsubsection{ Generic case }

Apart from the two extreme cases just discussed, one expects that 
the ground state wavefunction is characterized by a non-trivial multifractal spectrum.
For the pure case where $J_i=J$ and $h_i=h$, exact expressions
for the generalized dimension $D_q$ as a function of the control parameter
$K=\frac{J}{h}$ exist 
 in the limit $q=\pm \infty$ \cite{jms2009,jms2010,jms2011,atas_short,atas_long}
\begin{eqnarray}
D^{pure}_{q=\pm \infty} (K) = \frac{1}{2} - \frac{1}{2 \pi \ln 2}
 \int_0^{\pi} du \ln \left[ 1 \pm \frac{K-\cos u}{\sqrt{1+K^2-2 K \cos u}}  \right]
\label{dqinftyexact}
\end{eqnarray}
and thus also for $q=1/2$ as a consequence of the duality relation 
\cite{jms2009,jms2010,jms2011,atas_short,atas_long,luitz_spectro}
\begin{eqnarray}
D^{pure}_{q=\frac{1}{2}} (K) = 1- D^{pure}_{q=+\infty} \left(\frac{1}{K}\right)
\label{duality}
\end{eqnarray}
These functions are continuous in the control parameter $K$, but 
the derivatives are singular at the critical point $K_c=1$
\begin{eqnarray}
\partial_K D^{pure}_{q=\pm \infty} (K) && =  - \frac{1}{2 \pi \ln 2}
 \int_0^{\pi} du \left[ \pm \frac{1}{\sqrt{1+K^2-2 K \cos u}}
-\frac{K-\cos u}{1+K^2-2 K \cos u}    \right]
\nonumber \\
&& =  \mp \frac{1}{ \pi (\ln 2) (1+K) }{\cal K} \left(\frac{4 K}{(1+K)^2}  \right)
+ \frac{1}{ 2 K \ln 2} {\rm sgn} (K-1)
\label{dqinftyexactderi}
\end{eqnarray}
where ${\cal K}(m)\equiv \int_0^{\frac{\pi}{2}} \frac{d \theta}{\sqrt{1-m \sin^2 \theta}}$ is the elliptic integral of the first kind displaying the logarithmic singularity ${\cal K}(m) \simeq - \frac{1}{2} \ln (1-m)$ for $m \to 1^-$. As a consequence,
 the first term corresponds 
to a logarithmic divergence, whereas the second term corresponds to a discontinuity
\begin{eqnarray}
\partial_K D^{pure}_{q=\pm \infty} (K) && \opsimeq_{ \vert K -1 \vert \ll 1}
  \mp \frac{1}{ 2 \pi (\ln 2)  } \ln \vert K-1 \vert 
+ \frac{1}{ 2  \ln 2} {\rm sgn} (K-1)
\label{dqinftyexactderising}
\end{eqnarray}

Other values of $q$, especially the Shannon value $q=1$,
 have been studied numerically \cite{jms2009,jms2010,jms2011,atas_short,atas_long}.
In the following sections, we describe how the Fernandez-Pacheco real-space renormalization
allows to derive a simple approximation for the multifractal properties,
both for the pure and for the random chain.

\section{ Analysis via real-space renormalization }

\label{sec_rg}

The pure quantum Ising chain is the basic model to study quantum phase transitions at zero-temperature \cite{sachdev} : it is in the same universality class as the exactly solved 
2D classical Ising model. 
The random quantum Ising chain has been solved by Daniel Fisher \cite{fisher}  via the asymptotically exact strong disorder renormalization procedure : the transition is governed by an Infinite Disorder Fixed Point and presents unconventional scaling laws with respect to the pure case.

From the point of view of Block-Renormalization for quantum models, 
there exists a simple self-dual procedure introduced by Fernandez-Pacheco
\cite{pacheco,igloiSD},
which is able to reproduce the exact critical point $(J/h)_c=1$ and the exact correlation length exponent $\nu^{pure}=1$ of the pure chain \cite{pfeuty},
even if it is not able to reproduce the exact magnetic exponent \cite{pacheco}.
Its application to the disordered chain 
 also reproduces correctly the location of the critical point, the activated exponent $\psi=1/2$
and the two correlation length exponents \cite{nishiRandom,us_pacheco},
even if it is not able to reproduce the properties dominated by rare events 
\cite{fisher}.
In this section, we describe how it leads to simple renormalization rules for the 
Shannon and R\'enyi entropies of the ground state wavefunction.

\subsection{ Reminder on the Fernandez-Pacheco self-dual real space renormalization  }

The goal is to replace each block of two spins
$(\sigma_{2i-1};\sigma_{2i})$ by a single renormalized spin $\sigma_{R(2i)}$.
The idea of Fernandez-Pacheco \cite{pacheco}, 
written here for the case of arbitrary transverse fields and arbitrary ferromagnetic couplings \cite{nishiRandom,us_pacheco}, is 
to decompose the Hamiltonian of Eq. \ref{h1d} as the sum
$H=H^{intra}+H^{extra}$ with the following choice
\begin{eqnarray}
H_{intra} && =  \sum_{i} \left[  -    h_{2i-1} \sigma_{2i-1}^x 
- J_{2i-1} \sigma_{2i-1}^z  \sigma_{2i }^z   \right]
\nonumber \\
H_{extra} && =  \sum_{i} \left[  -    h_{2i} \sigma_{2i}^x 
- J_{2i} \sigma_{2i}^z  \sigma_{2i+1 }^z   \right]
\label{h1intra1d}
\end{eqnarray}

Since $H_{intra} $ is diagonal in the $\sigma^z$ basis for the even spins $\sigma_{2i } $, we have to diagonalize $H_{intra} $ independently in each subspace 
$(S_2,S_4,..,S_{2i},..)$. 
In addition, the odd spins $\sigma_{2i-1}$
are independent of each other and submitted to the one-spin Hamiltonian
\begin{eqnarray}
{\cal H}^{(S_{2i})}_{2i-1}= - h_{2i-1} \sigma_{2i-1}^x - J_{2i-1} S_{2i} \sigma_{2i-1}^z 
\label{hintraeffi}
\end{eqnarray}
The two eigenvalues are independent of the value of $S_{2i}=\pm 1$
\begin{eqnarray}
\lambda^{\pm}_{2i-1} = \pm \sqrt{ h_{2i-1}^2 +J_{2i-1}^2  }
\label{blocklambda}
\end{eqnarray}
with the following corresponding eigenvectors
\begin{eqnarray}
\vert \lambda^{-}_{2i-1}(S_{2i}) > && =
 \sqrt{ \frac{1 +  \frac{J_{2i-1}S_{2i}}{\sqrt{ h_{2i-1}^2 +J_{2i-1}^2  }} }{2} }
\vert S_{2i-1}=+1 >
+  \sqrt{ \frac{1 -  \frac{J_{2i-1}S_{2i}}{\sqrt{ h_{2i-1}^2 +J_{2i-1}^2  }} }{2} } \vert S_{2i-1}=-1 > 
\nonumber \\
\vert \lambda^{+}_{2i-1}(S_{2i}) > && = 
 - \sqrt{ \frac{1 -  \frac{J_{2i-1}S_{2i}}{\sqrt{ h_{2i-1}^2 +J_{2i-1}^2  }} }{2} }
\vert S_{2i-1}=+1 >
+  \sqrt{ \frac{1 +  \frac{J_{2i-1}S_{2i}}{\sqrt{ h_{2i-1}^2 +J_{2i-1}^2  }} }{2} } \vert S_{2i-1}=-1 > 
\label{blocklambda1eigen}
\end{eqnarray}

Projecting onto the ground state for the odd spins yields the renormalized
Hamiltonian for the even spins
\begin{eqnarray}
H^R && = \sum_{i} \left[  - h^R_{2i}  \sigma_{R(2i)}^x
  - J_{2i-2}^R    \sigma^z_{R(2i-2)} \sigma^z_{R(2i)} \right]
\label{h1dRfin}
\end{eqnarray}
with  the renormalized transverse fields 
\begin{eqnarray}
h^R_{2i} && =h_{2i} 
 \frac{h_{2i-1}}{\sqrt{h^2_{2i-1}+J^2_{2i-1} } }
\label{rgh1d}
\end{eqnarray}
and the renormalized couplings 
\begin{eqnarray}
 J^R_{2i-2} && = J_{2i-2} 
\frac{J_{2i-1}}{\sqrt{h^2_{2i-1}+J^2_{2i-1} } }
\label{rgj1d2x}
\end{eqnarray}

\subsection{ RG rules for the IPR and for the Shannon-R\'enyi entropies  }

From the point of view of the IPR of Eq. \ref{IPR}, the real-space renormalization yields
\begin{eqnarray}
Y_q(N) && = \sum_{S_1=\pm 1} \sum_{S_2=\pm 1} .. \sum_{S_N=\pm 1}
\vert \psi(S_1,S_2,...,S_N) \vert^{2q}
\nonumber \\
&&=  \sum_{S_2=\pm 1} \sum_{S_4=\pm 1} .. \sum_{S_N=\pm 1}
\vert \psi^R(S_2,S_4...,S_N) \vert^{2q}
\prod_{i=1}^{\frac{N}{2}} {\tilde y}_q(2i-1)
\label{IPRrg}
\end{eqnarray}
where ${\tilde y}_q(2i-1;2i)$ represents the IPR of the wavefunction 
$\lambda^{-}_{2i-1}(S_{2i}) $ (Eq. \ref{blocklambda1eigen})
of the odd spin $S_{2i-1}$ in the field of the even spin $S_{2i}$
\begin{eqnarray}
{\tilde y}_q(2i-1;2i) && \equiv
  \sum_{S_{2i-1}=\pm 1}  \vert < S_{2i-1}  \vert \lambda^{-}_{2i-1}(S_{2i}) >   \vert^{2q}
\nonumber \\
&& = \left[ \frac{1+\frac{J_{2i-1}}{\sqrt{h^2_{2i-1}+J^2_{2i-1} } }  }{2}  \right]^{q} 
+ \left[ \frac{1-\frac{J_{2i-1}}{\sqrt{h^2_{2i-1}+J^2_{2i-1} } }  }{2}  \right]^{q} 
\label{yq}
\end{eqnarray}
Since ${\tilde y}_q(2i-1;2i)$ 
is actually independent of the value $S_{2i}=\pm 1$, and only depends on the ratio
$K_{2i-1} \equiv \frac{J_{2i-1}}{h_{2i-1}} $, it is convenient 
to introduce the auxiliary function
\begin{eqnarray}
  y_q(K) \equiv \left[ \frac{1+\frac{K}{\sqrt{1+K^2} }  }{2}  \right]^{q} 
+  \left[ \frac{1-\frac{K}{\sqrt{1+K^2} }  }{2}  \right]^{q} 
\label{yqpurk}
\end{eqnarray}
Then Eq. \ref{IPRrg}
can be rewritten as
\begin{eqnarray}
Y_q(N) = \sum_{S_1=\pm 1} \sum_{S_2=\pm 1} .. \sum_{S_N=\pm 1}
\vert \psi(S_1,S_2,...,S_N) \vert^{2q}
= \left( \prod_{i=1}^{\frac{N}{2}} y_q(K_{2i-1}) \right) Y_q^R(\frac{N}{2})
\label{IPRrgfin}
\end{eqnarray}
where
\begin{eqnarray}
 Y_q^R(\frac{N}{2}) \equiv \sum_{S_2=\pm 1} \sum_{S_4=\pm 1} .. \sum_{S_N=\pm 1}
\vert \psi^R(S_2,S_4...,S_N) \vert^{2q}
\label{IPRduR}
\end{eqnarray}
is the IPR of the renormalized system of the $\frac{N}{2}$ even spins
with the renormalized parameters given in Eqs \ref{rgh1d} and \ref{rgj1d2x}.

The multiplicative RG rule of Eq. \ref{IPRrgfin} 
translates into the following additive RG rule
 for the Shannon-R\'enyi entropies of Eq. \ref{renyi}
\begin{eqnarray}
S_q(N) = \frac{\ln Y_q(N) }{1-q} 
= \sum_{i=1}^{\frac{N}{2}} \frac{\ln y_q(K_{2i-1})}{1-q} + S_q^R(\frac{N}{2})
\label{rglogIPR}
\end{eqnarray}
where $S_q^R(\frac{N}{2})=\frac{\ln Y_q^R(\frac{N}{2}) }{1-q}  $
is the entropy of the renormalized system of the $\frac{N}{2}$ even spins
with the renormalized parameters given in Eqs \ref{rgh1d} and \ref{rgj1d2x}.

In the following sections, we analyze the consequences of the RG rule of Eq. \ref{rglogIPR}
for the pure chain and for the random chain respectively.

\section { Multifractal spectrum for the pure case }

\label{sec_pure}

\subsection{ RG rules for the couplings \cite{pacheco}}

If the initial parameters are $(h,J)$ on the whole chain,
 the Fernandez-Pacheco renormalization yields
after one RG step the renormalized parameters (Eqs \ref{rgh1d} and \ref{rgj1d2x})
\begin{eqnarray}
h^R && = h  \frac{  h }{\sqrt{J^2 + h^2}}
\nonumber \\
J^R && = J  \frac{  J }{\sqrt{J^2 + h^2}}
\label{rgrulespur1d}
\end{eqnarray}
so that the ratio $K \equiv \frac{J}{h}$ evolves according to the simple rule
\begin{eqnarray}
K^R \equiv \frac{J^R}{h^R} =  K^2 \equiv \phi(K)
\label{krpur}
\end{eqnarray}
The disordered attractive fixed point $K=0$ and the 
ferromagnetic attractive fixed point $K \to +\infty$ are separated by
the unstable fixed point $K_c=1$ characterized by the correlation length
exponent $\nu=1$ obtained by $2^{\frac{1}{\nu}}=\phi'(K_c)=2 K_c=2$.
So  the location of the critical point $K_c=1$ and and
the correlation length exponent $\nu=1$ are in agreement with the exact solution
 \cite{pfeuty}.

\subsection{ RG rule for the Shannon-R\'enyi entropies }

The factors $y_q(K_{2i-1})$ in Eq. \ref{yq}
are all the same, and depend only on the control parameter $K=J/h$ 
so that the RG rule of Eq. \ref{rglogIPR} for the Shannon-R\'enyi entropies reads
\begin{eqnarray}
S_q(K;N)
= \frac{N}{2} \frac{\ln y_q(K)}{1-q} + S_q(K_R=K^2;\frac{N}{2})
\label{rglogIPRpure}
\end{eqnarray}
where the function $y_q(K)$ has been introduced in Eq. \ref{yqpurk}.
The iteration yields
\begin{eqnarray}
S_q(K;N)
= \frac{N}{2} \frac{\ln y_q(K)}{1-q}  
+ \frac{N}{4} \frac{\ln y_q(K^2)}{1-q}
+ \frac{N}{8} \frac{\ln y_q(K^4)}{1-q}
+...
+ \frac{N}{2^k} \frac{\ln y_q(K^{2^{k-1}})}{1-q}
+...
\label{rglogIPRpureiter}
\end{eqnarray}
where the $k^{th}$ term corresponds to the contribution of
the  $k^{th}$ RG step, where $\frac{N}{2^k}$ spins coupled via $K^{2^{k-1}} $ are eliminated,
and where $\frac{N}{2^k}$ spins survive with the renormalized coupling $K^{2^k}$.

\subsection{ Generalized fractal dimensions $D_q(K)$  }

From Eq. \ref{rglogIPRpureiter}, one obtains the generalized fractal dimensions $D_q(K)$
of Eq. \ref{renyi} as a series over the RG steps $k=1,2,..,+\infty$
\begin{eqnarray}
D_q(K) && = \lim_{N \to +\infty} \frac{ S_q(K; N)}{  N \ln 2}
=  \sum_{k=1}^{+\infty} \frac{d_q(K^{2^{k-1}} ) }{2^k } 
\label{dqK}
\end{eqnarray}
in terms of the auxiliary function $d_q(K)$ obtained from Eq. \ref{yqpurk}
\begin{eqnarray}
d_q(K) \equiv \frac{\ln y_q(K ) }{ (1-q) \ln 2} 
=\frac{\ln \left(  \left[ \frac{1+\frac{K}{\sqrt{1+K^2} }  }{2}  \right]^{q} 
+  \left[ \frac{1-\frac{K}{\sqrt{1+K^2} }  }{2}  \right]^{q} 
\right) }{ (1-q) \ln 2} 
\label{auxdqpurK}
\end{eqnarray}
that becomes for the Shannon value $q=1$
\begin{eqnarray}
d_{q=1}(K) && 
=-  \frac{\left[ \frac{1+\frac{K}{\sqrt{1+K^2} }  }{2} \right]
\ln  \left[ \frac{1+\frac{K}{\sqrt{1+K^2} }  }{2}  \right] 
+ \left[ \frac{1-\frac{K}{\sqrt{1+K^2} }  }{2} \right]
\ln \left[ \frac{1-\frac{K}{\sqrt{1+K^2} }  }{2}  \right] }{  \ln 2} 
\label{auxdq1purK}
\end{eqnarray}

As it should, one recovers at the paramagnetic fixed point $K=0$ where $d_q(K=0) =1$
the value of Eq. \ref{dqpara}
\begin{eqnarray}
D_q(K=0) =1
\label{rglogIPRzero}
\end{eqnarray}
and at the ferromagnetic fixed point $K=+\infty$ where $d_q(K=+\infty) =0$
the value of Eq. \ref{dqferro}
\begin{eqnarray}
D_q(K=+\infty) =0
\label{rglogIPRinfty}
\end{eqnarray}

\subsection{ Multifractal spectrum at the critical point $K_c=1$  }

\begin{figure}[htbp]
 \includegraphics[height=6cm]{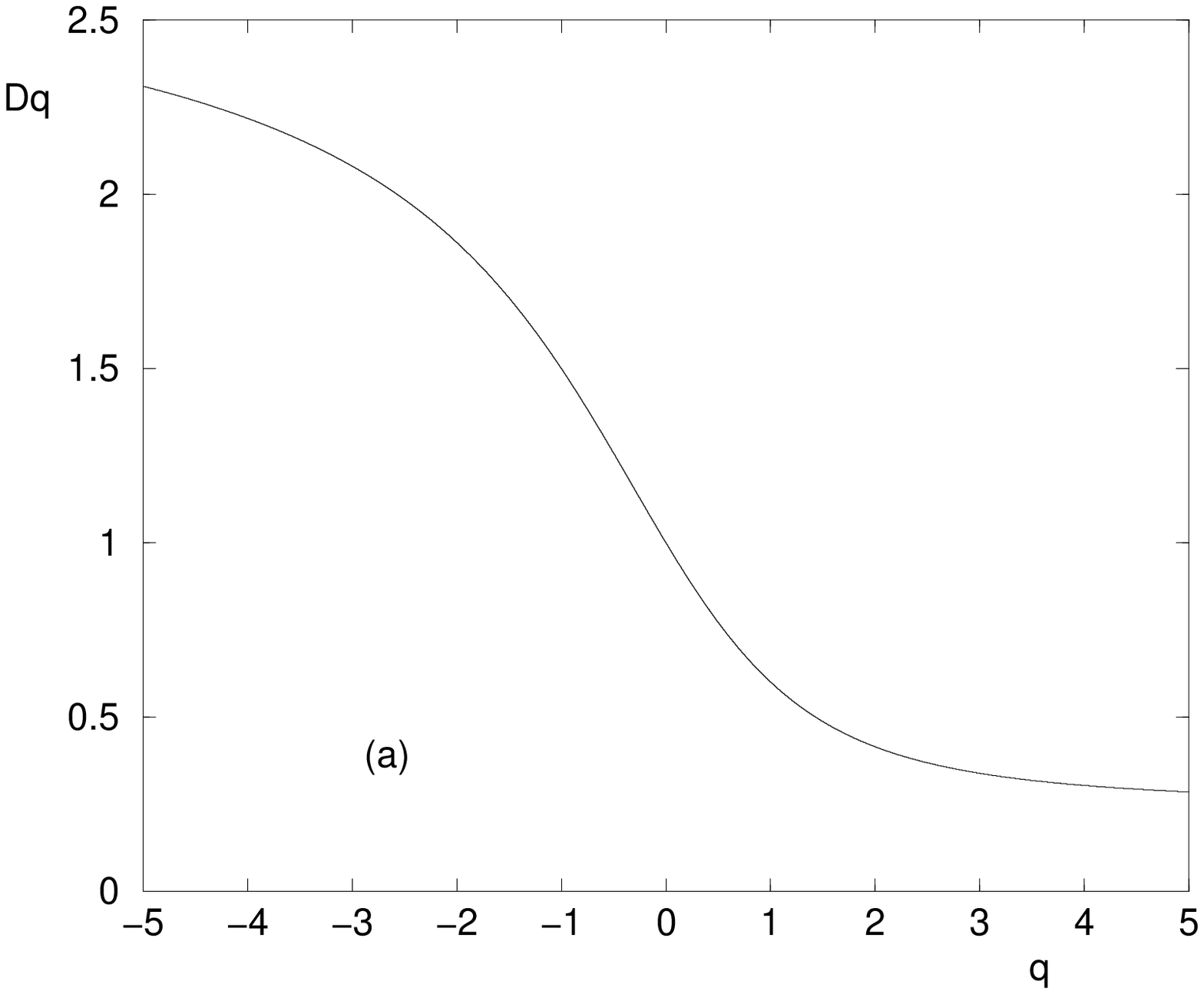}
\hspace{1cm}
 \includegraphics[height=6cm]{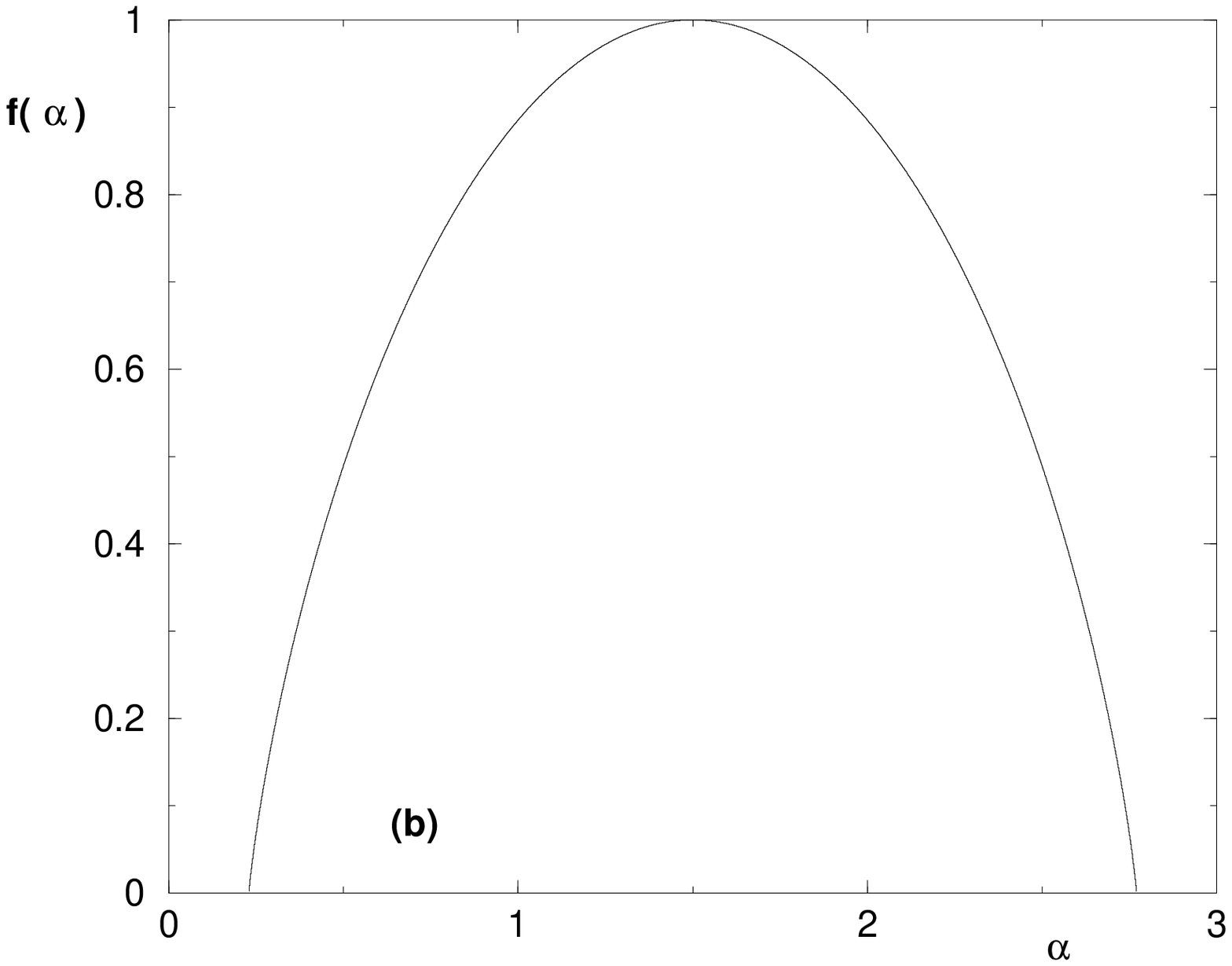}
\caption{ Pure quantum Ising chain at the critical point $K_c=1$ \\
(a) Generalized fractal dimension $D_q(K_c=1)$ as a function of the R\'enyi index $q$\\
(b) Corresponding multifractal spectrum $f(\alpha)$ as a function of $\alpha$.  }
\label{figmultifcriti}
\end{figure}

At the critical point $K_c=1$, the generalized fractal dimensions $D_q(K_c=1) $
of Eq. \ref{dqK}
simply coincide with the auxiliary function of Eq. \ref{auxdqpurK}
\begin{eqnarray}
D_q(K_c=1) &&= d_q(1)
 = \frac{ \ln \left( \left[ \frac{1-\frac{1}{\sqrt{2} }  }{2}  \right]^{q} 
+  \left[ \frac{1+\frac{1}{\sqrt{2} }  }{2}  \right]^{q}   \right) }{(1-q) \ln 2}
\label{dqcriti}
\end{eqnarray}
This function is plotted on Fig. \ref{figmultifcriti} (a) for 
the R\'enyi index in the interval $q \in [-5,+5]$.
Let us quote some special values
\begin{eqnarray}
D_{q \to 0}(K=1) &&  = 1- \frac{q}{2}+O(q^2)
\nonumber \\
D_{q=1/2}(K=1) &&  = \frac{ \ln \left(1+\frac{1}{\sqrt 2} \right) }{\ln 2} = 0.771553
\nonumber \\
D_{q=1}(K=1) &&  = \frac{ 
-\left( \frac{1-\frac{1}{\sqrt{2} }  }{2} \right)
\ln \left( \frac{1-\frac{1}{\sqrt{2} }  }{2}\right) 
-\left( \frac{1+\frac{1}{\sqrt{2} }  }{2} \right)
\ln \left( \frac{1+\frac{1}{\sqrt{2} }  }{2}\right) 
}{\ln 2} =  0.600876
\nonumber \\
D_{q=2}(K=1) &&  = \frac{\ln 4 - \ln 3}{\ln 2}=0.415037
\nonumber \\
D_{q=+\infty} (K=1)&&  =- \frac{\ln  \left[ \frac{1+\frac{1}{\sqrt{2} }  }{2}  \right]}{\ln 2} = 0.228447
\label{dqcritiplusinfinity}
\end{eqnarray}
in order to compare them with the explicit exact values for $q=1/2$
 and $q=+\infty$ \cite{atas_short,atas_long}
 and the numerical values for $q=1$ and $q=2$ \cite{jms2010}
\begin{eqnarray}
D^{exact}_{q=1/2} (K=1)&& =\frac{ 2 Catalan}{\pi \ln 2} = 0.841267..
\nonumber \\
D^{num}_{q=1} (K=1)&&  =\frac{0.42327...}{\ln 2 } = 0.61065...
\nonumber \\
D^{num}_{q=2} (K=1)&&  =\frac{0.21380...}{\ln 2 } = 0.30845...
\nonumber \\
D^{exact}_{q=+\infty} (K=1)&& = 1- \frac{ 2 Catalan}{\pi \ln 2} =0.158733..
\label{dqcritipluq1}
\end{eqnarray}
Our conclusion is thus that the RG yields a very good approximated value for the Shannon value $q=1$, whereas the approximation is clearly worse for the other R\'enyi indices $q \ne 1$. 
This seems to indicate that the Fernandez-Pacheco RG procedure is well suited for the initial Hamiltonian corresponding to $q=1$, whereas it
should probably be modified for $q \ne 1$ in order to take into account the deformation of the measure with respect to the initial quantum problem corresponding to $q=1$.
Indeed from the point of view of the equivalent 2D classical Ising model, the case where $M=2q$ is an integer
corresponds to the 'Ising book' (see Fig. 4 of \cite{jms2010}) where $M=2q$ half-planes are glued together.
This clearly changes the measure of the spin distribution and should be taken into account within the RG procedure
in order to obtain better values for $q \ne 1$.

On Fig. \ref{figmultifcriti} (b), the corresponding  multifractal spectrum $f_{K_c=1}(\alpha)$
is plotted as a function of $\alpha$ (see section \ref{sec_falpha} for a reminder on its meaning) from the parametric form of Eqs \ref{alphaq} and \ref{falphaq}
\begin{eqnarray}
   \alpha_q(K_c=1) && = 
 - \frac{1 }{  \ln 2} \left(
\frac{\left[ \frac{1+\frac{1}{\sqrt{2} }  }{2}  \right]^{q}
 \ln \left[ \frac{1+\frac{1}{\sqrt{2} }  }{2}  \right] 
+  \left[ \frac{1-\frac{1}{\sqrt{2} }  }{2}  \right]^{q} 
\ln \left[ \frac{1-\frac{1}{\sqrt{2} }  }{2}  \right] }
{\left[ \frac{1+\frac{1}{\sqrt{2} }  }{2}  \right]^{q} 
+  \left[ \frac{1-\frac{1}{\sqrt{2} }  }{2}  \right]^{q} } \right)
\nonumber \\
   f_{K_c=1}(\alpha_q) 
&& =  q \alpha_q(K_c=1) +
 \frac{ \ln \left( \left[ \frac{1-\frac{1}{\sqrt{2} }  }{2}  \right]^{q} 
+  \left[ \frac{1+\frac{1}{\sqrt{2} }  }{2}  \right]^{q}   \right) }{ \ln 2}
\label{falphaqrescriti}
\end{eqnarray}
In particular for $q=0$, the typical exponent of Eq. \ref{alphazero}
takes the simple value
\begin{eqnarray}
   \alpha_0(K_c=1) && = \frac{3}{2}
\label{alphazerocriti}
\end{eqnarray}

\subsection{ Singularities at the critical point $K_c=1$  }

\begin{figure}[htbp]
 \includegraphics[height=6cm]{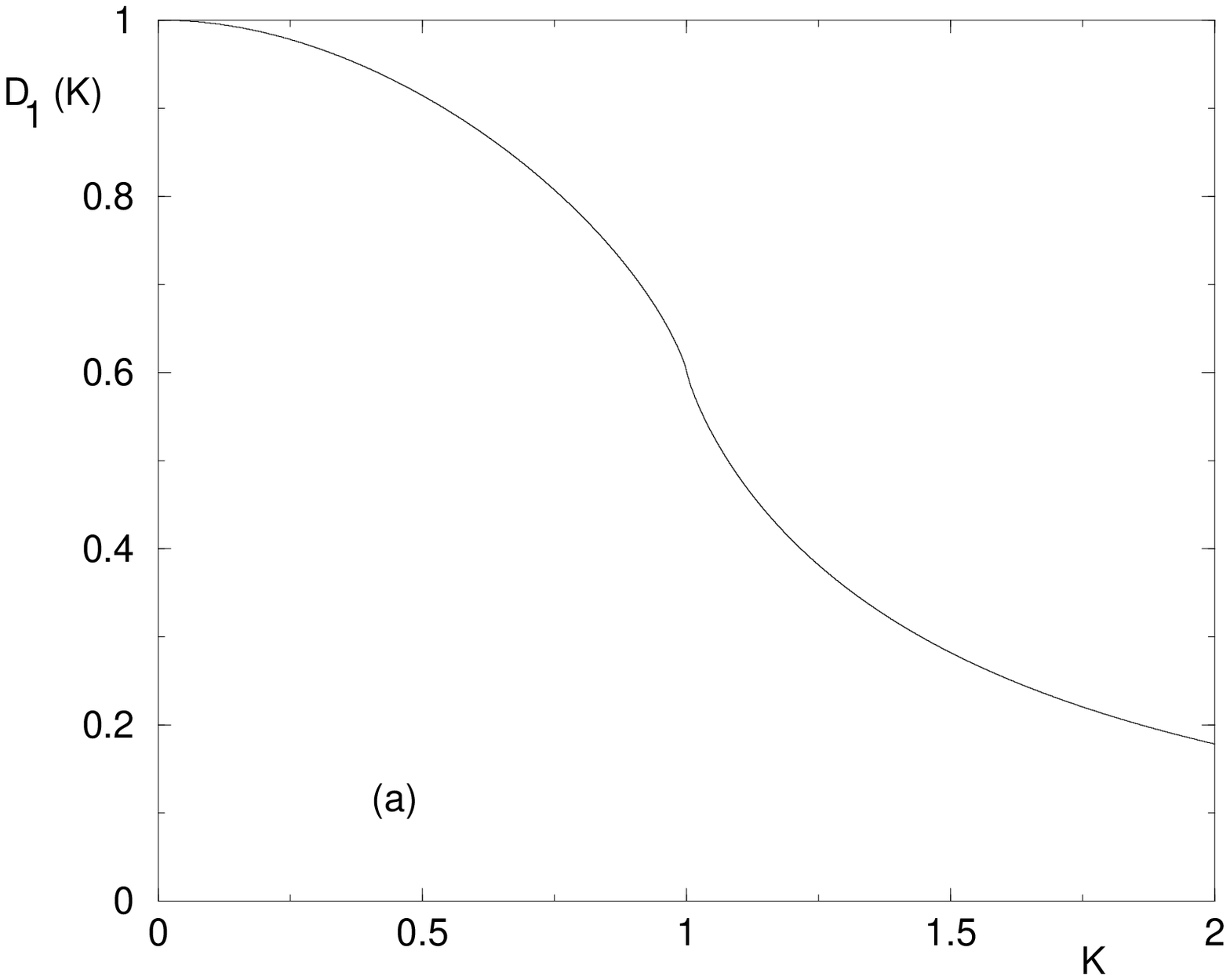}
\hspace{1cm}
 \includegraphics[height=6cm]{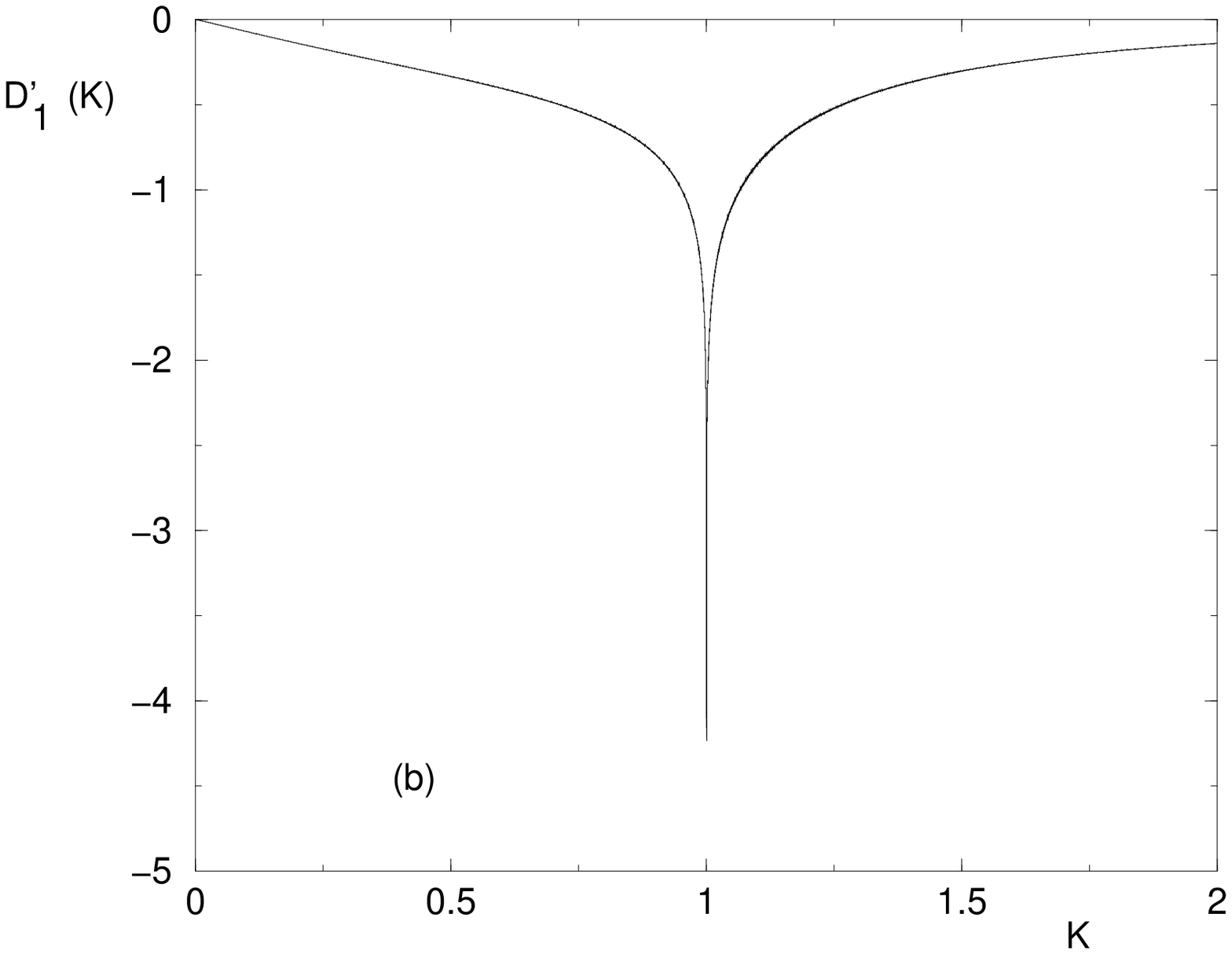}
\caption{ Pure quantum Ising chain at the Shannon value $q=1$ \\
(a) The generalized fractal dimension $D_1(K)$ is continuous
 a function of the control parameter $K=\frac{J}{h}$\\
(b) The derivative $\partial_K D_1(K) $ as a function of the control parameter $K=\frac{J}{h}$ displays a logarithmic singularity at the critical point $K_c=1$.  }
\label{figdq1k}
\end{figure}

The values $D_q(K=1)$ are not universal, but the derivatives of $D_q(K)$ 
with respect to the control parameter $K$
are expected to display universal singularities at the critical point.
The first derivative of Eq. \ref{dqK}
\begin{eqnarray}
 D_q'  (K) && = \frac{1}{2 K }
  \sum_{k=1}^{+\infty}  K^{2^{k-1} }d_q'(K^{2^{k-1}} )
\label{dqKderi}
\end{eqnarray}
clearly diverges at $K_c=1$.
Since this divergence is due to the region of large RG-step $k$,
we may evaluate the form of the singularity by replacing the series over the integer $k$
by the corresponding integral over the real variable $k$,
 and then perform a change of variable
from $k$ to $K_R=K^{2^{k-1}} $
\begin{eqnarray}
 D_q'  (K) && \opsimeq_{ \vert K-1 \vert \ll 1}  \frac{1}{2  }
  \int_{1}^{+\infty} dk  K^{2^{k-1} }d_q'(K^{2^{k-1}} )
\nonumber \\
&& \opsimeq_{ \vert K-1 \vert \ll 1}  \frac{1}{2  }
  \int_{K}^{+\infty} \frac{dK_R }{  (\ln K_R)  (\ln 2)}   d_q'(K_R )
\label{dqKderising}
\end{eqnarray}
i.e. the second derivative displays a pole singularity
\begin{eqnarray}
 D_q''  (K) && \opsimeq_{ \vert K-1 \vert \ll 1} - \frac{  d_q'(K )  }{2(\ln 2) (\ln K)  }
\opsimeq_{ \vert K-1 \vert \ll 1} - \frac{  d_q'(1 )  }{2 (\ln 2) ( K-1)  }+...
\label{dqKderi2sing}
\end{eqnarray}
The singularity of the first derivative is thus given by the following logarithmic divergence
\begin{eqnarray}
 D_q'  (K) && 
\opsimeq_{ \vert K-1 \vert \ll 1} - \frac{  d_q'(1 )  }{2 (\ln 2)   } \ln \vert K-1 \vert +...
\label{dqKderi1sing}
\end{eqnarray}
This behavior is in agreement with the logarithmic singularity
of Eq. \ref{dqinftyexactderising} concerning the limit $q \to \pm \infty$.

Finally, the function itself is continuous with the following logarithmic singularity
\begin{eqnarray}
 D_q  (K) && \opsimeq_{ \vert K-1 \vert \ll 1} D_q(1) - \frac{  d_q'(1 )  }{2 (\ln 2)   }
(K-1)  \ln \vert K-1 \vert +...
\label{dqKsing}
\end{eqnarray}

For the Shannon index $q=1$, the generalized fractal dimension $D_1(K)$ 
and its derivative  $\partial_K D_1(K) $ are plotted as a function of the 
 control parameter $K=\frac{J}{h}$ on Fig. \ref{figdq1k}.
The corresponding coefficient of the logarithmic singularity 
on these figures is
\begin{eqnarray}
 - \frac{  d_q'(1 )  }{2 (\ln 2)   } = - \frac{{\rm arccoth }(\sqrt{2})}{4 {\sqrt 2} (\ln 2)^2} = -0.32429
\label{dqKsingcoefq1}
\end{eqnarray}

\subsection{ Duality functional relation  }

As a final remark, let us mention that the RG result of Eq. \ref{dqK}
satisfies the exact duality relation quoted in Eq. \ref{duality}
Indeed for the index value $q=1/2$, the auxiliary function of Eq. \ref{auxdqpurK}
reads
\begin{eqnarray}
d_{q=\frac{1}{2}}(K)
&& = \frac{\ln \left(  \left[ \frac{1+\frac{K}{\sqrt{1+K^2} }  }{2}  \right]^{\frac{1}{2}} 
+  \left[ \frac{1-\frac{K}{\sqrt{1+K^2} }  }{2}  \right]^{\frac{1}{2}} 
\right)^2 }{ \ln 2} 
=  \frac{\ln \left(  \frac{1+\frac{K}{\sqrt{1+K^2} }  }{2} + \frac{1-\frac{K}{\sqrt{1+K^2} }  }{2}
+ 2 \sqrt{ \frac{1-\frac{K^2}{1+K^2 }  }{4} }
\right) }{ \ln 2} 
\nonumber \\
&& =  \frac{\ln \left( 1 +   \frac{1}{\sqrt{1+K^2}}
\right) }{ \ln 2} 
\label{auxdqpurKdemi}
\end{eqnarray}
whereas for the index value $q=+\infty$, it reads
\begin{eqnarray}
d_{q=+\infty}(K) 
=- \frac{\ln \left(  \frac{1+\frac{K}{\sqrt{1+K^2} }  }{2} 
\right) }{  \ln 2} = 1 - d_{q=\frac{1}{2}}\left(\frac{1}{K}\right)
\label{auxdqpurKinfty}
\end{eqnarray}
so that the duality relation of Eq. \ref{duality} is satisfied term by term in the series representation of Eq. \ref{dqK}.

\section { Multifractal spectrum for the random chain }

\label{sec_random}

\subsection{ RG rules for the couplings \cite{nishiRandom,us_pacheco}}

In terms of the ratios
\begin{eqnarray}
K_i \equiv \frac{J_i}{h_{i+1}} 
\label{ki}
\end{eqnarray}
the RG rules of Eq. \ref{rgh1d} and \ref{rgj1d2x} read
\begin{eqnarray}
K^R_{2 i-2} && \equiv \frac{J^R_{2i-2}}{h^R_{2i}} 
 = \frac{J_{ 2i-2}  J_{2i-1}} { h_{2i-1} h_{2i} }  =K_{2i-2} K_{2i-1} 
\label{rgruleskr}
\end{eqnarray}
that generalize the pure RG rules of Eq. \ref{krpur}.

Since this corresponds to a simple addition in log-variables
\begin{eqnarray}
\ln K^R_{2i-2} =\ln K_{2i-2} + \ln K_{2i-1} 
\label{rgruleslogkr}
\end{eqnarray}
one obtains after $k$ RG steps in terms of the initial variables
\begin{eqnarray}
\ln K^{R^k} = \sum_{i=1}^{2^k} \ln K_{i-1} =
 \sum_{i=1}^{2^k} \left[ \ln J_{i-1}-\ln h_i   \right]
\label{inffp}
\end{eqnarray}
The Central Limit theorem yields the asymptotic behavior
\begin{eqnarray}
\ln K^{R^k} \opsimeq_{k \to + \infty} 2^k \left[ \overline{ \ln J_{i-1} -\ln h_i }  \right]
+ \sqrt{2^k} \sqrt{ \left[ Var[\ln  J_{i} ] + Var[\ln h_i ]   \right] } u
\label{inffpclt}
\end{eqnarray}
where $u$ is a Gaussian random variable of zero mean and of variance unity.
So for large RG step $k$, the probability 
distribution  $P_k(K_R)$ of the renormalized coupling $K_R=K^{R^k}$
is log-normal 
\begin{eqnarray}
 P_k(K_R) \opsimeq_{k \to +\infty} 
 \frac{1}{ K_R \sqrt{2 \pi V 2^{k} } } e^{- \frac{ (\ln K_R - 2^{k} \ln K)^2}{2 V 2^{k}} } 
\label{lognormal}
\end{eqnarray}
with the notations 
\begin{eqnarray}
 \ln K && \equiv \overline{ \ln J_{i-1} -\ln h_i }
\nonumber \\
 V && \equiv  Var[\ln  J_{i} ] + Var[\ln h_i ]
\label{lognormalnot}
\end{eqnarray}

The first term of Eq. \ref{inffpclt}
yields that the critical point corresponds to the condition
\begin{eqnarray}
\ln K_c \equiv \overline{\ln J_{i-1} -\ln h_i  }  =0
\label{criti1d}
\end{eqnarray}
and that the typical correlation length exponent with respect to the length scale $L=2^k$ is
\begin{eqnarray}
\nu_{typ}=1
\label{ntyp1d}
\end{eqnarray}
 Outside criticality, the competition between the first and the second term 
 of Eq. \ref{inffpclt} shows
that the finite-size correlation exponent is 
\begin{eqnarray}
\nu_{FS}=2
\label{nav1d}
\end{eqnarray}
At criticality where the first term vanishes, the second random
term of order $L^{1/2}$ corresponds to an Infinite Disorder Fixed Point of exponent
\begin{eqnarray}
\psi=\frac{1}{2}
\label{psi1d}
\end{eqnarray}

All these conclusions of Eqs \ref{criti1d}, \ref{ntyp1d}, \ref{nav1d}, \ref{psi1d}
obtained via the application of the Fernandez-Pacheco renormalization to
the random quantum Ising chain \cite{nishiRandom,us_pacheco}, are
in agreement with the Fisher Strong Disorder renormalization exact results \cite{fisher}.

\subsection{ RG rule for averaged Shannon-R\'enyi entropies }

Averaging Eq. \ref{rglogIPR} over the disorder yields the
following RG rule for averaged Shannon-R\'enyi entropies
\begin{eqnarray}
\overline{ S_q(N) } 
= \frac{N}{2} \frac{\overline{ \ln y_q(K_{2i-1}) }}{1-q} +\overline{ S_q^R(\frac{N}{2}) }
\label{rglogIPRav}
\end{eqnarray}
where $\overline{S_q^R(\frac{N}{2})}  $
is the averaged entropy of the renormalized system of the $\frac{N}{2}$ even spins
with the renormalized parameters given in Eq \ref{rgruleskr}.
Note that whenever multifractality appears in disordered systems
(see for instance the review \cite{mirlinrevue} on Anderson localization transitions), 
one should make the distinction between the 'averaged multifractal sprectrum'
based on the disorder-average of the IPR, and the 'typical multifractal sprectrum'
based on the disorder-average of the logarithm of the IPR.
Here we only consider the disorder-averaged entropies of Eq. \ref{rglogIPRav},
corresponding to the disorder-average of the logarithm of the IPR that define
 the typical multifractal spectrum.

\subsection{ Generalized fractal dimensions $D_q$ }

\begin{figure}[htbp]
 \includegraphics[height=6cm]{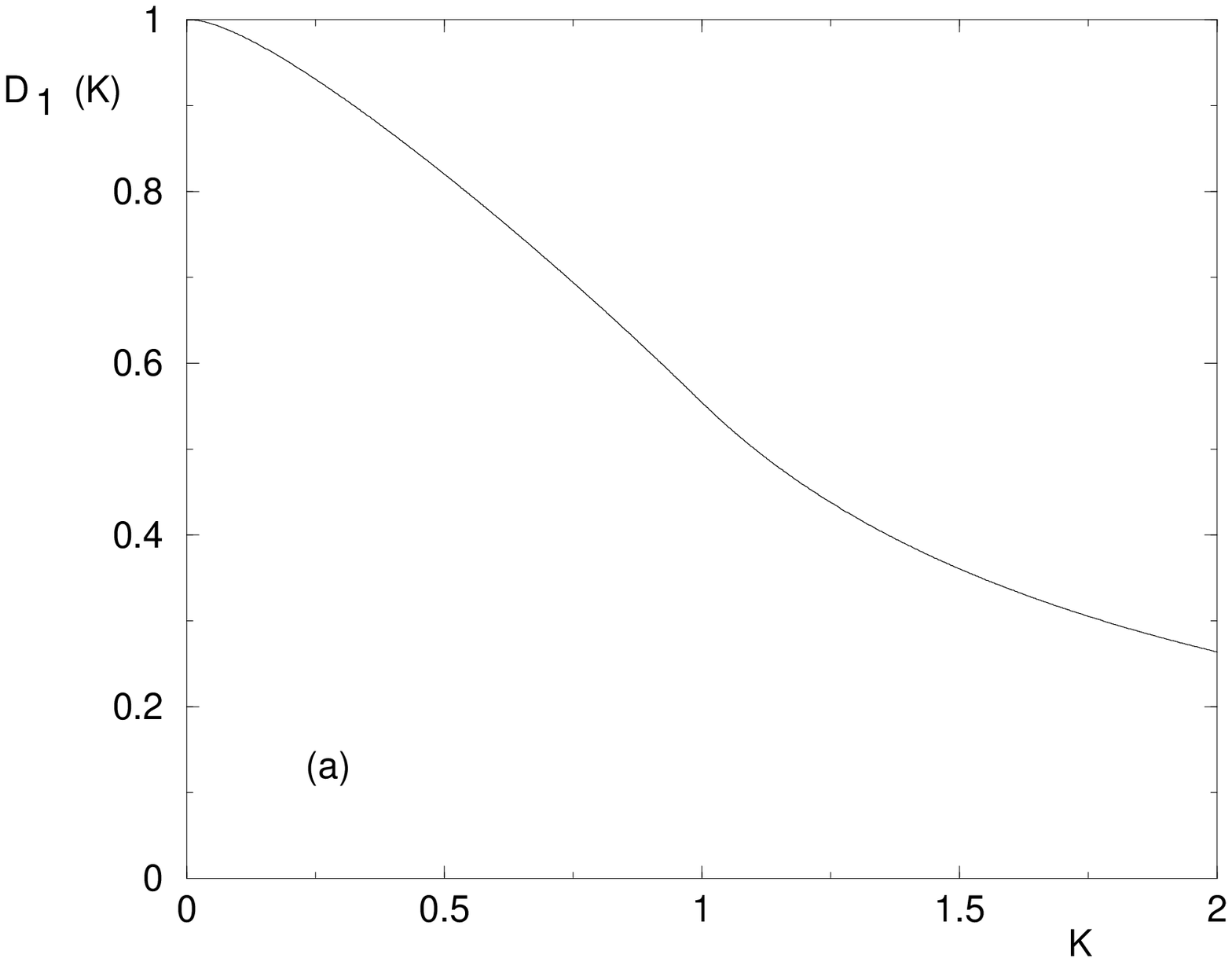}
\hspace{1cm}
 \includegraphics[height=6cm]{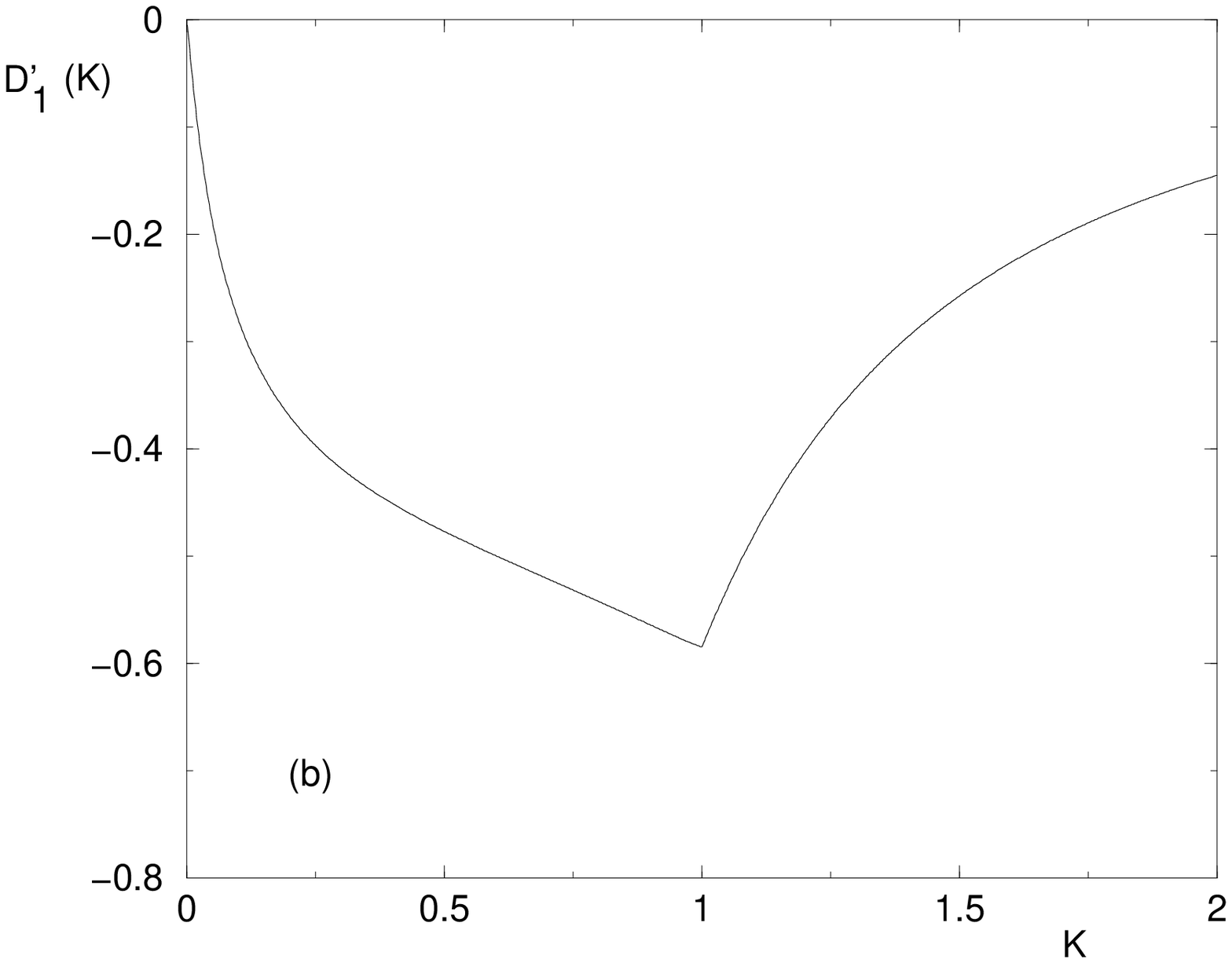}
\caption{ Random quantum Ising chain at the R\'enyi index $q=1$ \\
(a) Information dimension $D_1(K)$ as a function of the control parameter $K=e^{\overline{\ln J_i-\ln h_i}} $\\
(b) The derivative $\partial_K D_1(K) $ as a function of the control parameter $K$ displays
 a cusp singularity at the critical point $K_c=1$.  }
\label{figdq1krandom}
\end{figure}

So in terms of the auxiliary function $d_q(K)$ of Eq. \ref{auxdqpurK},
the iteration of Eq. \ref{rglogIPRav}
yields the generalized fractal dimension
\begin{eqnarray}
D_q && \equiv \lim_{N \to +\infty} \frac{\overline{ S_q( N) }}{  N \ln 2}
=  \sum_{k=1}^{+\infty} \int_0^{+\infty} dK_R P_{k-1}(K_R) \frac{ d_q(K_R ) }{2^k } 
\label{dqKrandom}
\end{eqnarray}
where $P_k(K_R)$ is the probability distribution of the renormalized variable $K_R$ after $k$ RG steps (Eq. \ref{inffp}).
Of course the values $D_q$ of Eq. \ref{dqKrandom} are non-universal as in the pure case, since they depend on the contribution of the first RG steps.
In particular they thus depend upon the initial disorder distribution $P_0(K)$.
On Fig. \ref{figdq1krandom}, we show $D_1(K)$ and its derivative $\partial_K D_1(K)$
as a function of the control parameter $K$ of Eq. \ref{lognormalnot}
for the special case where the initial disorder distribution $P_0(K)$ 
is itself given by the log-normal form of Eq. \ref{lognormal}
with the value $V=1$, so that Eq. \ref{lognormal} is valid for all RG steps $k$
\begin{eqnarray}
D_q (K) && 
=  \sum_{k=1}^{+\infty} \int_0^{+\infty} dK_R 
 \frac{dK_R   }{ K_R \sqrt{2 \pi V 2^{k-1} } } e^{- \frac{ (\ln K_R - 2^{k-1} \ln K)^2}{2 V 2^{k-1}} } 
 \frac{ d_q(K_R ) }{2^k } 
\nonumber \\
&& = 
\sum_{k=1}^{+\infty}\int_{-\infty}^{+\infty} 
 \frac{du}{  \sqrt{2 \pi } } e^{- \frac{ u^2}{2 } } 
 \frac{ d_q( K^{2^{k-1} } e^{ \sqrt{ V 2^{k-1} } u }  ) }{2^k } 
\label{dqKlognormal}
\end{eqnarray}
On Fig. \ref{figdq1krandom}, it is clear that $D_1(K)$ and its derivative
 $\partial_K D_1(K)$ are both continuous at the critical point
$K_c=1$, but that the derivative displays a cusp (see the difference with the corresponding Figure \ref{figdq1k} concerning the pure case).
To understand the origin of this singularity, we may take into account
the Infinite Disorder nature of the critical point and the Strong Disorder nature
of the critical region. More precisely at large RG step $k$,
 the renormalized coupling
\begin{eqnarray}
K_R \equiv  K^{2^{k-1} } e^{ \sqrt{ V 2^{k-1} } u } 
\label{dqKcrandom}
\end{eqnarray}
is either very small (if $2^{k-1} \ln K +\sqrt{ V 2^{k-1} } u <0  $)
or either very big (if $2^{k-1} \ln K +\sqrt{ V 2^{k-1} } u >0  $)
with the following leading contributions of the auxiliary function of Eq. \ref{auxdqpurK}
\begin{eqnarray}
d_q(K) && \opsimeq_{K \to 0} 1 +  O(K^2)
\nonumber \\
d_q(K) && \opsimeq_{K \to +\infty} 0 +O({\rm max} (\frac{1}{ K^2},\frac{1}{ K^{2q}}))
\label{lnyqpurkzero}
\end{eqnarray}
Let us thus make the approximation that the singular part is given 
in the critical region by
\begin{eqnarray}
D_q^{sing} (K) && 
\simeq  
\sum_{k=1}^{+\infty} \frac{ 1 }{2^k }  \int_{-\infty}^{- \frac{2^{k-1} \ln K}{\sqrt{ V 2^{k-1} }}} 
 \frac{du}{  \sqrt{2 \pi } } e^{- \frac{ u^2}{2 } } 
\label{dqKsingran}
\end{eqnarray}
Since the singularity is due to the region of large RG-step $k$,
we may evaluate the first derivative by replacing the series over the integer $k$
by the corresponding integral over the real variable $k$,
 and then perform a change of variable from $k$ to $x= \frac{ 2^{k-1} }{2 V } (\ln K)^2$
\begin{eqnarray}
\partial_K D_q^{sing} (K)  
&& \simeq  
- 2 \sum_{k=1}^{+\infty}    \frac{ 1 }{\sqrt{2 \pi V 2^{k-1} } K } 
 e^{- \frac{ 2^{k-1} }{2 V } (\ln K)^2 } 
\nonumber \\
&& \simeq  
- 2 \int_{1}^{+\infty} dk   \frac{ 1 }{\sqrt{2 \pi V 2^{k-1} } K } 
 e^{- \frac{ 2^{k-1} }{2 V } (\ln K)^2 } 
\nonumber \\
&& \simeq  
- \frac{ \vert \ln K \vert }{ (\ln 2) V \sqrt{ \pi    } K} \int_{\frac{ (\ln K)^2 }{2 V } }^{+\infty}
    \frac{ dx }{ x^{\frac{3}{2} }  } 
 e^{- x } 
\nonumber \\
&& \simeq  
- \frac{ 1 }{ (\ln 2) V \sqrt{ \pi    } K}
\left[ 2 \sqrt{2V}  -2 \sqrt{\pi}  \vert \ln K \vert
 + 2 \frac{ ( \ln K )^2 }{ \sqrt{2 V} } + O\left(  (\ln K)^2  \right) \right]
\label{dqKransingderi1}
\end{eqnarray}
is continuous at the critical point $K_c=1$, but displays a cusp as a consequence of
the second term containing the factor $\vert \ln K \vert \propto \vert K-1  \vert $
near criticality.

\subsection{ Analysis of the Shannon-R\'enyi entropies via Strong Disorder RG}

For the random quantum Ising chain, the strong disorder renormalization procedure
has been shown to be asymptotically exact by Daniel Fisher \cite{fisher}
so that it can be used to derive all universal critical properties
(see \cite{fisher} and the review \cite{review_strong}),
as well entanglement properties (see \cite{refael} and the review \cite{review_entang}).
It is thus interesting to consider its application to the Shannon-R\'enyi entropies,
even if only the singularities will be meaningful, whereas the 
non-universal values dominated by the first RG steps will 
of course not be reproduced faithfully.

In the Strong Disorder RG procedure \cite{fisher,review_strong},
the strongest term $\Omega$ among the ferromagnetic couplings $J_i$ and the transverse fields
 $h_i$
\begin{eqnarray}
\Omega={\rm max} (J_i,h_i)
\label{omega}
\end{eqnarray}
 is decimated iteratively. If the strongest term is a coupling $J_i$,
the two spins linked by this coupling are coupled ferromagnetically
to form a new renormalized spin 
(the two kept states are $\vert ++>$ and $\vert -->$).
So the contribution of this
decimation to
the Shannon-R\'enyi entropies is actually zero
\begin{eqnarray}
S^{Jdecim}_q(N) && = S^{R}_q(N-1)
\label{strongrgj}
\end{eqnarray}
If the strongest term is a transverse field
 $h_i$, the corresponding spin is in the eigenstate of $\sigma^x$,
so that the contribution of this
decimation to
the Shannon-R\'enyi entropies is the maximal value $\ln 2$
\begin{eqnarray}
S_q^{hdecim}(N) && = S^{R}_q(N-1)+(\ln 2) 
\label{strongrg}
\end{eqnarray}
As a consequence, the generalized fractal dimension is simply 
\begin{eqnarray}
D_q = \frac{\overline{S_q(N)}}{N \ln 2}  && = \frac{ {\cal N}_h }{{\cal N}_h+{\cal N}_J}
\label{dqstrongrg}
\end{eqnarray}
where ${\cal N}_h$ is the total number of $h$-decimations along the RG flow,
and ${\cal N}_J $ is the total number of $J$-decimations along the RG flow.

We refer to Refs \cite{fisher,review_strong} for the renormalization rules
 and for the asymptotic probability distributions of the renormalized couplings and of the renormalized transverse fields as a function of the RG scale $\Gamma=-\ln \Omega$
and of the usual control parameter $\delta$ \cite{fisher,review_strong}
related to the previous notations $(K,V)$ of Eq. \ref{lognormalnot}
\begin{eqnarray}
\delta \equiv \frac{\overline{ \ln J_{i-1} -\ln h_i }}
{ Var[\ln  J_{i} ] + Var[\ln h_i ]} = \frac{ \ln K}{V}
\label{delta}
\end{eqnarray}
Here we only need the probability $P_0(\Gamma) d\Gamma $ 
to decimate a coupling in the interval $[\Gamma,\Gamma+d \Gamma]$
and the probability $R_0(\Gamma)d\Gamma$
to decimate a transverse field in the interval $[\Gamma,\Gamma+d \Gamma]$,
that read \cite{fisher,review_strong}
\begin{eqnarray}
P_0(\Gamma)= \frac{2 \delta }{1- e^{2 \delta \Gamma}}
\nonumber \\
R_0(\Gamma)= \frac{2 \delta }{ e^{2 \delta \Gamma}-1}
\label{jhdecim}
\end{eqnarray}
In particular, the density $n_{\delta}(\Gamma)$ of surviving spins at scale $\Gamma$ evolves according to \cite{fisher}
\begin{eqnarray}
\partial_{\Gamma} n_{\delta}(\Gamma) = - \left[P_0(\Gamma)+R_0(\Gamma)  \right]  n_{\delta}(\Gamma)
\label{dndelta}
\end{eqnarray}
and the solution starting from the initial condition $n_{\delta}(\Gamma_0) =1$
at the initial scale $\Gamma_0$ reads \cite{fisher}
\begin{eqnarray}
n_{\delta}(\Gamma) = \frac{\sinh^2 \delta \Gamma_0}{\sinh^2 \delta \Gamma}
\label{ndelta}
\end{eqnarray}
The fraction of $h$-decimations is then given in the thermodynamic limit by
\begin{eqnarray}
D_q(\delta) && = \frac{ {\cal N}_h }{{\cal N}_h+{\cal N}_J }
 = \frac{ {\cal N}_h }{N} = \int_{\Gamma_0}^{+\infty} d \Gamma  R_0(\Gamma) n_{\delta}(\Gamma)
 =\sinh^2 \delta \Gamma_0
 \int_{\Gamma_0}^{+\infty} d \Gamma  \frac{ \delta e^{- \delta \Gamma} }{\sinh^3 \delta \Gamma}
\nonumber \\
&& = \sinh^2 \delta \Gamma_0 \left[ \frac{ \frac{1}{2} e^{-2 \delta \Gamma}-1}{\sinh^2 \delta \Gamma}   \right]_{\Gamma=\Gamma_0}^{\Gamma=+\infty}
\label{fractionhdecim}
\end{eqnarray}

At criticality $\delta=0$, the fraction of $h$-decimation is of course 
\begin{eqnarray}
D_q(\delta=0) && = \frac{1}{2}
\label{fractionhdecimcriti}
\end{eqnarray}
by symmetry.
In the ordered phase $\delta>0$, it is given by
\begin{eqnarray}
D_q(\delta>0) && = 1- \frac{1}{2} e^{- 2 \delta \Gamma_0}
\label{fractionhdecimor}
\end{eqnarray}
whereas in the disordered phase $\delta<0$ it reads
\begin{eqnarray}
D_q(\delta<0) && = \frac{1}{2} e^{ 2 \delta \Gamma_0}
\label{fractionhdecimdis}
\end{eqnarray}
In particular the derivative
\begin{eqnarray}
\partial_{\delta} D_q(\delta) && =\Gamma_0 e^{- 2 \vert \delta \vert \Gamma_0}
\label{fractionhdecideri}
\end{eqnarray}
is continuous but displays a cusp at criticality $\delta=0$,
in agreement with the result \ref{dqKransingderi1} of the Fernandez-Pacheco renormalization.

\section{ Conclusion }

\label{sec_conclusion}

 In this paper, we have studied via real space renormalization the 
 Shannon and the R\'enyi entropies of the ground state wavefunction in the pure and in the random quantum Ising chain. Our main conclusion is that the leading extensive term presents the following singularities at the quantum phase transition : the derivative with respect to the control parameter is logarithmically divergent in the pure case, and  displays a cusp singularity in the random case. This cusp singularity for the random case has been also confirmed via the Strong Disorder Renormalization approach. Although the logarithmic divergence for the pure quantum chain is reminiscent of the logarithmic divergence of the specific heat of the 2D classical Ising model, the cusp singularity found for the random quantum chain is different from the specific heat singularity of the corresponding 2D McCoy-Wu model
\cite{mccoywu}.

We should stress that the leading extensive terms of the Shannon and the R\'enyi entropies do not depend on the boundary conditions. As other extensive thermodynamic observables, their values are dominated by the small scales in the bulk and are thus non-universal, 
whereas their singularities at the critical point are dominated by the large scales and are thus universal. The subleading non-extensive terms of the Shannon and the R\'enyi entropies for finite chains of size $N$ are of course also very interesting. In the pure chain, these subleading contributions are universal, but depend upon the boundary conditions as follows :

(i) for the periodic chain, the subleading term is of order $O(1)$ and is non-trivial only at the critical point $K_c=1$ and for the Shannon value $q=1$
\cite{jms2009,jms2010,jms2011,grassberger,luitz_short,luitz_spectro}, whereas  
the direct application of the real space RG described in the text would
produce a non-trivial $O(1)$ term only at the critical point $K_c=1$,
but for all value of the R\'enyi index $q$. This suggests,
as already mentioned after Eq. \ref{dqcritipluq1},
  that the RG procedure probably needs to be improved for $q \ne 1$,
to take into account the deformation of the measure with respect to the initial quantum problem corresponding to $q=1$.

(ii) for the open chain, the subleading term at criticality is logarithmic
and $q$-dependent \cite{moore,luitz_spectro,jms2014,alcaraz},
as predicted from the presence of 'corners' in the 
 Conformal Field Theory of the replicated associated statistical physics model.
It would be interesting to better understand the origin of these logarithmic terms 
directly in the framework of the quantum Ising chain, and to reproduce them by some appropriate real-space RG procedure.

\section*{Acknowledgments}

It is a pleasure to thank Gr\'egoire Misguich and Fabien Alet
 for very useful discussions.

\end{document}